\documentclass[twocolumn,amsmath,amssymb,floatfix,nature,shownopacs,footinbib,superscriptaddress]{revtex4-1}
\usepackage{amsmath,amssymb,natbib,bm,graphicx,url,epsf,color}
\usepackage[british]{babel}
\usepackage{wasysym}
\usepackage{MnSymbol}
\setcitestyle{super} 

\begin{document}

\title{Two-channel Kondo effect and renormalization flow with macroscopic quantum charge states}

\author{Z. Iftikhar}
\affiliation{CNRS, Laboratoire de Photonique et de Nanostructures
(LPN), 91460 Marcoussis, France}
\author{S. Jezouin}
\affiliation{CNRS, Laboratoire de Photonique et de Nanostructures
(LPN), 91460 Marcoussis, France}
\author{A. Anthore}
\affiliation{CNRS, Laboratoire de Photonique et de Nanostructures
(LPN), 91460 Marcoussis, France}
\affiliation{Univ Paris Diderot, Sorbonne Paris Cit\'e, LPN, 91460 Marcoussis, France}
\author{U. Gennser}
\affiliation{CNRS, Laboratoire de Photonique et de Nanostructures
(LPN), 91460 Marcoussis, France}
\author{F.D. Parmentier}
\affiliation{CNRS, Laboratoire de Photonique et de Nanostructures
(LPN), 91460 Marcoussis, France}
\author{A. Cavanna}
\affiliation{CNRS, Laboratoire de Photonique et de Nanostructures
(LPN), 91460 Marcoussis, France}
\author{F. Pierre\thanks{frederic.pierre@lpn.cnrs.fr}}
\email[e-mail: ]{frederic.pierre@lpn.cnrs.fr}
\affiliation{CNRS, Laboratoire de Photonique et de Nanostructures
(LPN), 91460 Marcoussis, France}


\maketitle

{\sffamily
Many-body correlations and macroscopic quantum behaviors are fascinating condensed matter problems.
A powerful test-bed for the many-body concepts and methods is the Kondo model \cite{Vojta2006,Bulla2008} which entails the coupling of a quantum impurity to a continuum of states.
It is central in highly correlated systems \cite{Hewson1997,Cox1998,Dzero2010} and can be explored with tunable nanostructures \cite{Goldhaber-Gordon1998a,Cronenwett1998,Sasaki2004,Potok2007}.
Although Kondo physics is usually associated with the hybridization of itinerant electrons with microscopic magnetic moments \cite{Kondo1964}, theory predicts that it can arise whenever degenerate quantum states are coupled to a continuum \cite{Matveev1991,Cox1998,Yi1998,LeHur2012,Goldstein2013}. 
Here we demonstrate the previously elusive `charge' Kondo effect in a hybrid metal-semiconductor implementation of a single-electron transistor, with a quantum pseudospin-$1/2$ constituted by two degenerate macroscopic charge states of a metallic island \cite{Glazman1990,Matveev1991,Matveev1995,Furusaki1995b,Zarand2000,LeHur2002,Lebanon2003}.
In contrast to other Kondo nanostructures, each conduction channel connecting the island to an electrode constitutes a distinct and fully tunable Kondo channel \cite{Matveev1991}, thereby providing an unprecedented access to the two-channel Kondo effect and a clear path to multi-channel Kondo physics \cite{Nozieres1980,Cox1998,Pustilnik2004,Vojta2006}.
Using a weakly coupled probe, we reveal the renormalization flow, as temperature is reduced, of two Kondo channels competing to screen the charge pseudospin.
This provides a direct view of how the predicted quantum phase transition develops across the symmetric quantum critical point \cite{Nozieres1980,Cox1998}.
Detuning the pseudospin away from degeneracy, we demonstrate, on a fully characterized device, quantitative agreement with the predictions for the finite-temperature crossover from quantum criticality \cite{Furusaki1995b}.
}

\begin{figure}[!htbp]
\renewcommand{\figurename}{\textbf{Figure}}
\renewcommand{\thefigure}{\textbf{\arabic{figure}}}
\centering\includegraphics[width=1\columnwidth]{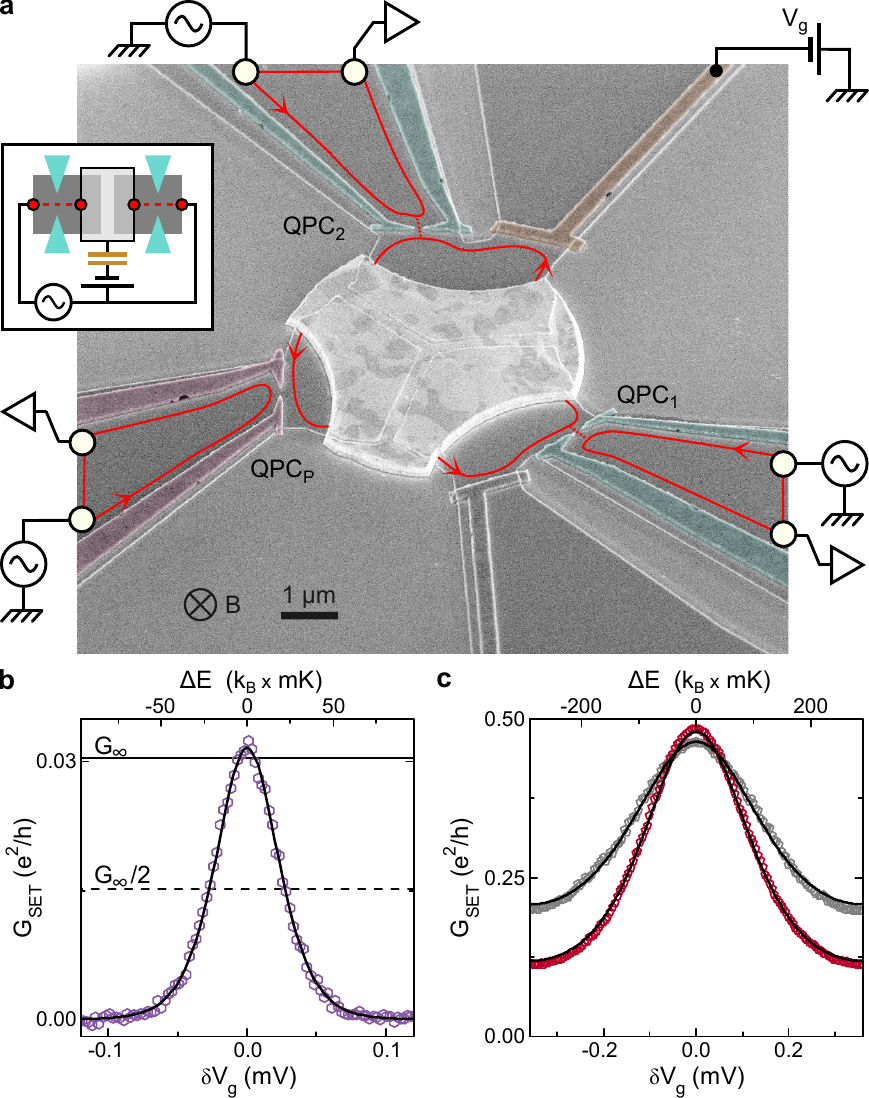}
\caption{
\small
\textbf{Hybrid metal-semiconductor single-electron transistor.} 
\textbf{a}, Colorized picture of the sample (schematic in inset) constituted of a central metallic island (bright) connected to large electrodes (white circles) through the quantum point contacts QPC$_{1,2}$ formed in a buried 2D electron gas (darker gray).
The lateral continuous gates and QPC$_p$ are used, respectively, to characterize the `intrinsic' and `in-situ' (renormalized) conductances of QPC$_{1,2}$.
The magnetic field $\mathrm{B}\simeq 3.9$~T corresponds to the integer quantum Hall regime, with the current propagating along spin-polarized edge channels (red lines) in the direction indicated by arrows. 
\textbf{b}, \textbf{c}, Kondo renormalized Coulomb peaks. Measured SET conductance (symbols) versus gate voltage $V_g$ (pseudospin energy splitting $\Delta E$), for symmetric QPC$_{1,2}$ set to $\tau_{1,2}\simeq 0.06$ (b) at $T\simeq 11.5$~mK or $\tau_{1,2}\simeq 0.93$ (c) at $T\simeq 11.5$~mK (red) and 22~mK (gray).
Continuous lines are theoretical predictions (see main text, Methods). 
The agreement data-theory in \textbf{c} establishes the predictions for the crossover from quantum critical behavior as a function of $\Delta E$ (main text).
\normalsize
}
\label{fig-sample}
\end{figure}

In previous experimental investigations, the Kondo quantum impurity was of microscopic nature and mostly associated with spin\cite{Goldhaber-Gordon1998a,Cronenwett1998,Nygard2000,Park2002,Liang2002,Potok2007}, orbital\cite{Sasaki2004,JarilloHerrero2005}, or possibly structural degrees of freedom\cite{Ralph1994,Cox1998}.
In the `charge' Kondo effect\cite{Matveev1991,Matveev1995,Furusaki1995b}, it is a pseudospin-$1/2$ constituted of two degenerate states of a macroscopic quantum variable, the electrical charge of a metallic island comprising several billions of electrons. 
The role of the electrons' spin ($\uparrow\downarrow$) in the original spin Kondo problem\cite{Kondo1964} is played by the electrons' location, in the island ($\uparrow$) or elsewhere ($\downarrow$). 
Accordingly, the charge pseudospin flips when electrons are transferred in and out of the island.
The Kondo channels, each coupling the Kondo impurity (pseudo)spin with a distinct electron continuum, directly equate with the different electrical conduction channels connected to the island (distinguishing between those associated with different values of the real electron spin).
In contrast, the electrical channels in previous Kondo nanostructures normally merged into a single Kondo channel (except in the ingenious implementation of ref.~\citenum{Potok2007}), due to cooperative spin-flip processes involving charge transfers between continuums.
Furthermore, the charge pseudospin energy splitting, adjusted by detuning the island from degeneracy with a gate voltage, is fully equivalent to the Zeeman splitting of a magnetic Kondo impurity.
Finally, of practical importance, the macroscopic charge pseudospin allows for large channel distances, and thereby enables full and independent control as well as the in-situ characterization of every Kondo parameter, giving access to direct comparisons with theory.

Here, we investigate a nanostructure designed to display the two-channel `charge' Kondo effect \cite{Matveev1991,Matveev1995,Furusaki1995b}. 
The device (Fig.~1a) is a hybrid metal-semiconductor single-electron transistor (SET) with additional characterization probes. 
It essentially consists of a central metallic island (bright), with a continuous electronic density of states, connected to large electrodes through two quantum point contacts (QPC$_{1,2}$), each tuned to a single conduction channel.
The QPCs are formed in a Ga(Al)As two-dimensional electron gas (2DEG) by field effect using split gates.
The 2DEG is further confined by etching (to the darker gray areas), and electrically connected to the metallic island by thermal annealing.
The lateral continuous gates are used to extract the `intrinsic' transmission probabilities $\tau_{1,2}$ characterizing QPC$_{1,2}$, respectively, by short-circuiting the central island. 
The capacitively coupled gate voltage $V_g$ controls the energy difference between the island charge states.
When set to weak coupling, QPC$_p$ gives us access separately to the `in-situ' conductances $G_{1,2}$ of QPC$_{1,2}$, respectively.
Except when specifically indicated, QPC$_p$ is disconnected.

The experiment is performed down to an electronic temperature $T\simeq 11.5$~mK (Methods), in a perpendicular magnetic field $\mathrm{B}\simeq 3.9$~T that breaks the spin degeneracy and corresponds to the integer quantum Hall effect at filling factor $2$. 
In this regime, the current flows along two (spin-polarized) chiral edge channels.
Red lines in Fig.~1a represent the outer channel, closest to the edge, with the propagation direction indicated by arrows.
It is partially transmitted across QPC$_{1,2}$, whereas the inner channel (not shown) is fully reflected and can be ignored.

\begin{figure}[!htbp]
\renewcommand{\figurename}{\textbf{Figure}}
\renewcommand{\thefigure}{\textbf{\arabic{figure}}}
\centering\includegraphics[width=0.9\columnwidth]{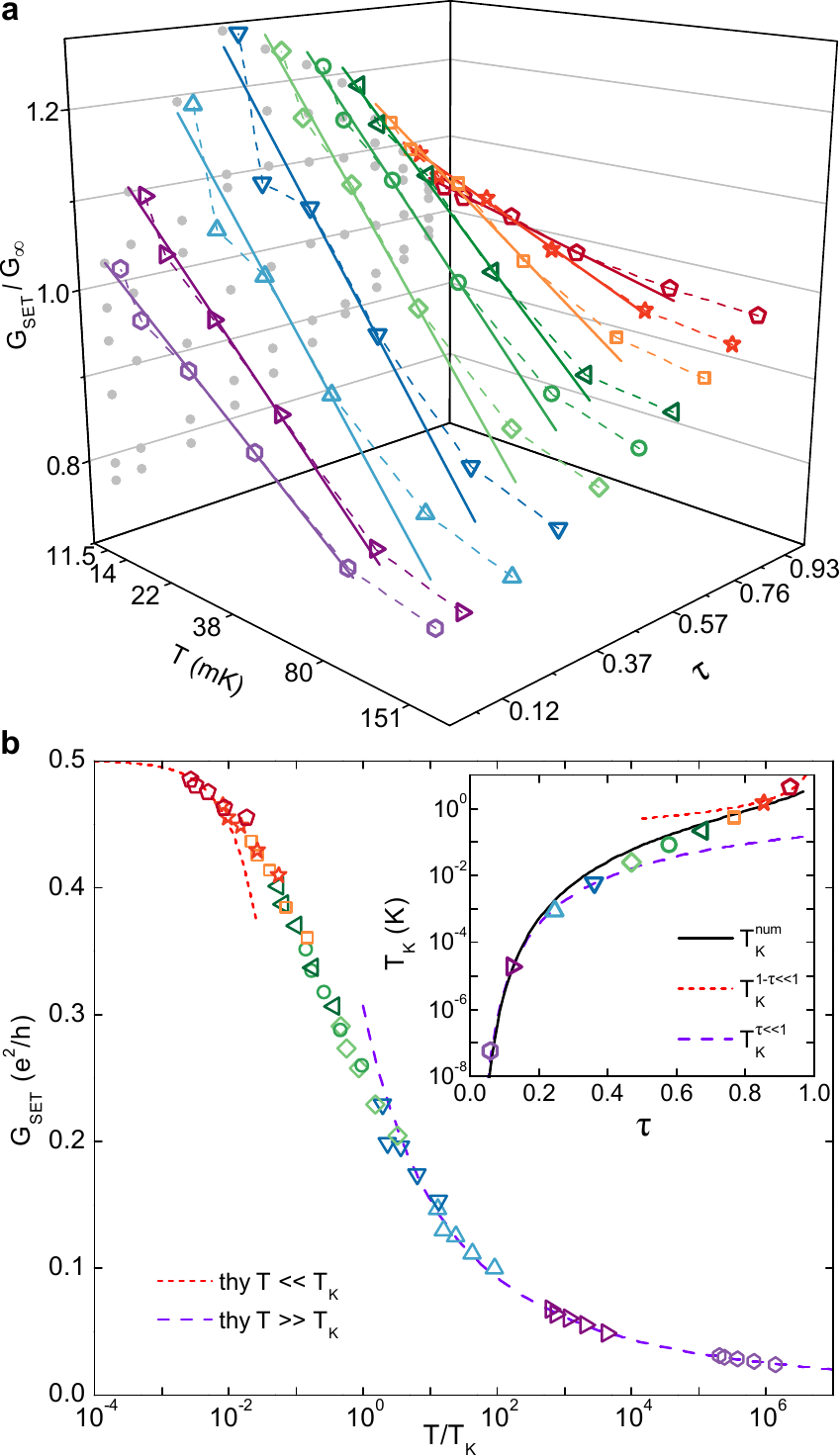}
\caption{\small
\textbf{Observation of the `charge' Kondo effect.} 
\textbf{a}, The normalized SET conductance $G_\mathrm{SET}/G_\infty$, at charge degeneracy ($\delta V_g=0$) and for symmetric QPC$_{1,2}$, is plotted as symbols versus the temperature on a log scale for different values of $\tau \equiv \tau_1 \simeq \tau_2$.
Continuous straight lines are guides to the eye proportional to $\log (T)$.
The grey dots are the orthogonal projections of the different temperature measurements onto the plane $(\tau,G_\mathrm{SET}/G_\infty)$.
\textbf{b}, The data in \textbf{a} at $T\leq 80$~mK, rescaled in temperature into a universal conductance curve (symbols).
The violet dashed line displays the theoretical (thy) $T\gg T_K$ prediction $G_\mathrm{SET}\propto \log^{-2}(T/\alpha T_K)$.
The red short-dashed line displays the $T\ll T_K$ prediction $e^2/2h-G_\mathrm{SET}\propto T/T_K$ (Methods). 
Inset, the extracted scaling parameter $T_K(\tau)$ (symbols) is compared to theoretical predictions (see Eqs.~\ref{eqTKtunnel} and \ref{eqTKnum} for the definitions of the Kondo temperatures $T_K^{\tau\ll1}$ and $T_K^\mathrm{num}$, and Methods for $T_K^{1-\tau\ll1}$).
\normalsize}
\label{fig-chargeKondo}
\end{figure}

We now review the main requirements for mapping the physics of this device to the two-channel Kondo (2CK) problem. 
Firstly, the typical electronic level spacing $\delta$ in the metallic island should be much smaller than the thermal energy: $\delta\ll k_BT$, with $k_B$ the Boltzmann constant\cite{Matveev1995,Furusaki1995b}.
We estimate $\delta\approx k_B\times0.2~\mu$K (Methods), nearly five orders of magnitude smaller than $k_BT$.
Secondly, the charging energy $E_C=e^2/2C$, with $e$ the electron charge and $C$ the overall island geometrical capacitance, should be larger than $k_BT$ to reduce the accessible charge states to a pseudospin-$1/2$.
We obtain from standard Coulomb diamond analysis $E_C\simeq k_B\times290~\mathrm{mK}$ (Methods). 
Thirdly, the metallic island should be in nearly perfect contact with the 2DEG, in particular to avoid resonances involving the 2DEG-metal interface. 
We find that the outer edge channel is fully transmitted into the metallic island, with a reflection probability smaller than $0.05$\% (Methods). 
Finally, QPC$_{1,2}$ should implement point-like contacts, with a small energy dependence of $\tau_{1,2}$. 
For the experimental set points, we find using the lateral characterization gates that $\tau_{1,2}$ increase monotonically with energy by at most 11\% up to $2 E_C$ (Methods). 
Together, the last two tests rule out any resonant effects.

From the influence of the charge states' energy splitting on conductance, we observe first indications of 2CK effects and establish that the measurements are performed in a regime where this physics is expected.
The measured conductance $G_\mathrm{SET}$ of the QPC$_1$-island-QPC$_2$ SET is shown as symbols versus gate voltage for symmetric QPCs set in the tunnel and weak-backscattering regimes, $\tau\equiv \tau_1\simeq\tau_2\simeq 0.06$ ($T\simeq 11.5~$mK) and $0.93$ ($T\simeq 11.5$ and 22~mK), in Fig.~1b,c respectively.
The conductance exhibits periodic peaks located at successive charge degeneracy points (one full period $\Delta \simeq 0.72$~mV is shown in Fig.~1c, Methods).

The tunnel data in Fig.~1b are compared with the prediction for incoherent sequential tunneling events \cite{Beenakker1991}
\begin{equation}
G_\mathrm{SET}=\frac{G_\infty}{2}\frac{2 E_C (\delta V_g/\Delta) /k_B T}{\sinh(2 E_C (\delta V_g/\Delta)/ k_B T)},
\label{eqseq}
\end{equation}
with $G_\infty=(e^2/h)/(\tau_1^{-1}+\tau_2^{-1})$ the `classical' SET conductance and $h$ the Plank constant.
We find that the data (symbols) can be accurately reproduced with a fit temperature of $10$~mK (continuous line), slightly smaller than but compatible with $T\simeq 11.5 \pm 1.5$~mK. 
However, the maximum peak conductance is much higher than the standard prediction $G_\infty/2$ (dashed line), and $G_\infty$ was left as a free fit parameter. 
Such an increase is expected from the Kondo renormalization of the conductance, even for relatively low characteristic Kondo temperature scales $T_K \ll T$. 
In this limit, Eq.~\ref{eqseq} is predicted to provide a good approximation when substituting $G_\infty$ by $\sim \log^{-2} (T/\alpha T_K)$, with $\alpha$ a numerical factor \cite{Furusaki1995b} (Methods). 
Assuming $\tau\ll 1$, the Kondo temperature reads \cite{Furusaki1995b}:
\begin{equation}
T_K^{\tau\ll 1} \sim (E_C/k_B) \exp (-\pi^2/\sqrt{4\tau}).
\label{eqTKtunnel}
\end{equation}

In the opposite limit of weak-backscattering ($1-\tau\ll1$), the 2CK physics is expected to be well developed. We find that the $\tau\simeq 0.93$ data (Fig.~1c, symbols) are accurately reproduced, quantitatively and without fit parameters, by the predictions (lines) from the theoretical framework where the Kondo mapping is established \cite{Furusaki1995b} (Methods).

With these indications of 2CK effects, we now provide direct experimental evidence of Kondo physics from the temperature dependence $G_\mathrm{SET}(T)$ at the charge degeneracy point, with QPC$_{1,2}$ remaining symmetric. 

In standard metallic SETs, with many opaque conduction channels, the peak conductance monotonically decreases from its high temperature classical value $G_\infty$ as the temperature is reduced \cite{Joyez1997}. 
In stark contrast, we find that $G_\mathrm{SET}(\delta V_g=0)$ increases as the temperature is reduced and, at $T\simeq 11.5$~mK, always exceeds the classical conductance $G_\infty$, by up to nearly 30\% (Fig.~2a). 
Note that the separately characterized intrinsic energy dependencies of $\tau_{1,2}$ correspond to an opposite decrease of $G_\mathrm{SET}$ smaller than 1\% for $T\lesssim 80$~mK.
Remarkably, the conductance increase is logarithmic in $T$ (continuous lines, for $T \lesssim 80~\mathrm{mK}$), which is a typical signature of the Kondo effect. 

A characteristic of Kondo systems, arising from renormalization group physics \cite{Bulla2008}, is that they follow universal scaling laws. 
We demonstrate that the conductance data at $T\leq 80$~mK can be rescaled into a single curve $G_\mathrm{SET}(T,\tau)=G_\mathrm{SET}(T/T_K)$, and that the extracted $T_K(\tau)$ agrees with the theoretical prediction for the Kondo temperature (Fig.~2b).
The simple rescaling procedure (Methods) relies on $G_\mathrm{SET}$ overlaps for different $\tau$, and on the prediction $G_\mathrm{SET}(T/T_K\gg 1)\propto \log^{-2} (T/\alpha T_K)$ (violet dashed line).
The $T\ll T_K$ prediction $e^2/2h-G_\mathrm{SET}\propto T/T_K$ is displayed as a red short-dashed line\cite{Furusaki1995b} (Methods).
The experimental scaling law covers an unprecedented range of $T/T_K$ and most of $G_\mathrm{SET}\in [0,0.5]e^2/h$, thanks to the fully and independently tunable $\tau$. 
Given the important $G_\mathrm{SET}$ overlaps, involving up to three successive values of $\tau$, the rescaling accuracy provides a stringent test of the universal scaling law hypothesis.  
This conclusion is further established by confronting extracted $T_K(\tau)$ (symbols in inset) with the predictions derived at $1-\tau\ll 1$ (Methods, red short-dashed line), $\tau\ll 1$ (Eq.~\ref{eqTKtunnel}, violet dashed line) and its generalization tested numerically (continuous line) \cite{Lebanon2003} 
\begin{eqnarray}
T_K^\mathrm{num} \sim (E_C/k_B) t\rho \exp(-\pi/(4t\rho)),
\label{eqTKnum}
\end{eqnarray}
where $\pi t\rho \simeq \sqrt{2(1-\sqrt{1-\tau})/\tau-1}$. 
Adjusting the unknown theoretical prefactor to match $T_K(\tau\ll 1)$ or $T_K(1-\tau\ll 1)$, we find an overall agreement over the whole range $\tau \in [0,1]$.

\begin{figure}[!htbp]
\renewcommand{\figurename}{\textbf{Figure}}
\renewcommand{\thefigure}{\textbf{\arabic{figure}}}
\centering\includegraphics [width=1\columnwidth]{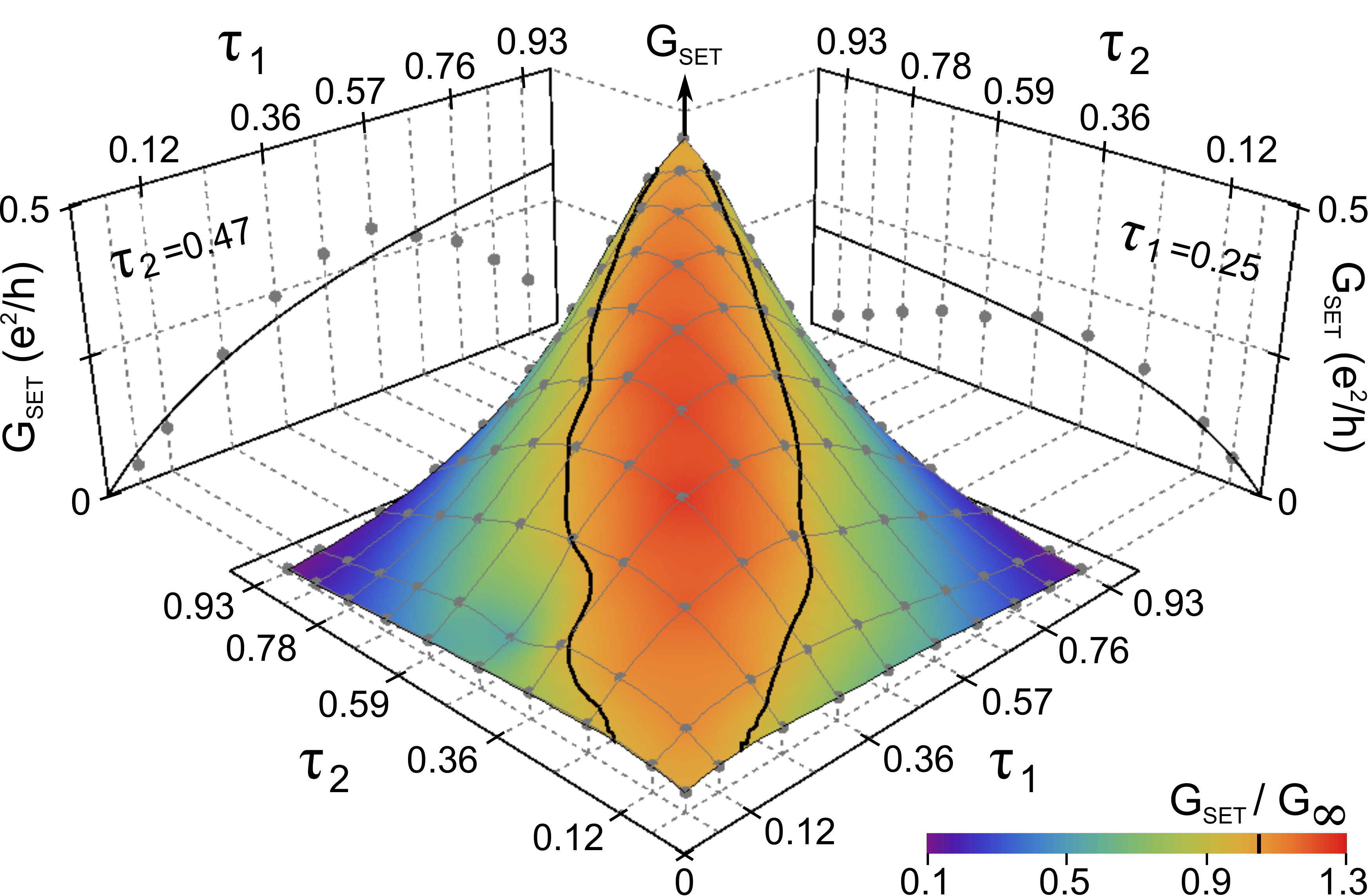}
\caption{\small
\textbf{Interplay of two Kondo channels revealed by tuning the asymmetry.} 
Main plot, the SET conductance at charge degeneracy and $T=11.5$~mK is displayed (symbols) versus QPC$_{1,2}$ `intrinsic' transmission probabilities $\tau_{1,2}$. 
Narrow gray lines connect data points with the setting of one QPC fixed while the other is changed.
The color code represents $G_\mathrm{SET}/G_\infty$ (black lines indicate $G_\mathrm{SET}=G_\infty$).
Lateral panels represent the same data (symbols) for a fixed value of $\tau_2\simeq 0.47$ (left) or $\tau_1\simeq 0.25$ (right), together with $G_\infty$ (continuous line). 
\normalsize}
\label{fig-asymGset}
\end{figure}

With the `charge' Kondo effect established, we turn to exploring the 2CK physics, which originates from the channels' competition to screen the (pseudo)spin-$1/2$.
Two symmetric Kondo channels are expected to flow, as $T \rightarrow 0$, toward the so-called\cite{Furusaki1995b} strong-coupling fixed point characterized in the `charge' Kondo implementation by two ballistic conduction channels ($G_{1,2}\rightarrow e^2/h$).
In contrast to the one-channel Kondo (1CK) effect, this produces an over-screening of the pseudospin and, consequently, a non-Fermi liquid 2CK state with collective low-energy excitations \cite{Nozieres1980}. 
The 2CK state is predicted to be unstable with an energy splitting of the pseudospin and with channel asymmetry, resulting in a $T=0$ quantum phase transition \cite{Nozieres1980}. 
Indeed, in the presence of an asymmetry the most strongly coupled Kondo channel takes over, fully screening the pseudospin-$1/2$ at low temperatures and thereby hiding (decoupling) it from the other channel (see ref.~\citenum{Potok2007} for first evidence of such a decoupling with a specific spin Kondo nanostructure\cite{Oreg2003}). 
From the quantum phase transition perspective, the 2CK non-Fermi liquid character appears as a general consequence of the divergent correlations near the quantum critical point (symmetric and at degeneracy) \cite{Vojta2006}. 
At finite $T \lesssim T_K$, the quantum critical (non-Fermi liquid) behavior is preserved for a range of channel asymmetries and pseudospin energy splittings, which narrows down as $T$ is reduced.
Consequently, a non-Fermi to Fermi liquid crossover takes place \cite{Nozieres1980,Cox1998,Pustilnik2004}.

The data in symmetric QPC configurations (Fig.~1b,c, Fig.~2b) already reveal information on the 2CK physics.
First, the experimental scaling law (Fig.~2b) shows that two symmetric Kondo channels flow monotonically toward the expected strong-coupling fixed point ($2G_\mathrm{SET}\simeq G_1 \simeq G_2 \rightarrow e^2/h$ as $T \rightarrow 0$). 
Note that for $N\geq 3$ Kondo channels, the predicted symmetric fixed point is different \cite{Nozieres1980,Furusaki1995b,Yi1998}.
In particular, for the ($N=3$)-channel `charge' Kondo effect, the in-situ conductance of each of the three (symmetric) QPCs is expected to flow toward $2\sin^2(\pi/5) e^2/h\simeq0.69e^2/h$ as $T \rightarrow 0$, see ref.~\citenum{Yi1998}.
Second, the crossover from quantum critical to Fermi liquid behavior with the pseudospin energy splitting $\Delta E=2E_C\delta V_g/\Delta$ is explored in Fig.~1c.
Starting from a well-developed 2CK state at $\delta V_g=0$ ($T/T_K\approx 0.003$ and $0.005$), the SET conductance progressively moves away from the strong coupling fixed point $e^2/2h$ and, at sufficiently large $\Delta E$, decreases as the temperature is reduced from 22 to $11.5$~mK.
Remarkably, the demonstrated agreement data-theory for arbitrary $\delta V_g$ validates, quantitatively, the theoretical description of the crossover \cite{Furusaki1995b}.
In particular, the crossover energy scale (such that $G_\mathrm{SET}=0.5\times e^2/2h$) increases with $T$, closely following the generic expectation $k_BT_K\sqrt{T/T_K}$ (Methods; we are preparing a thorough study of the crossover).

We provide a first evidence of the channels' competition by exploring the effect of QPC asymmetry on $G_\mathrm{SET}$. 
Symbols in Fig.~3 represent $G_\mathrm{SET}(\tau_1,\tau_2)$ measured at $T\simeq 11.5$~mK at the charge degeneracy point, while the color code corresponds to the ratio $G_\mathrm{SET}/G_\infty$. 
Note that it is only for nearly symmetric QPC$_{1,2}$ that $G_\mathrm{SET}$ exceeds the `classical' value $G_\infty$ (black continuous lines).
The stronger $G_\mathrm{SET}$ renormalization for symmetric QPCs indicates that they influence each other.
Strikingly, $G_\mathrm{SET}$ exhibits a maximum and then decreases as the transmission probability of one QPC is continuously increased with the other fixed (symbols in lateral panels). 
This non-monotonic behavior demonstrates that the two QPCs are not independently renormalized, and validates expectations for two competing Kondo channels in series \cite{Furusaki1995b,Pustilnik2004}.

\begin{figure}[!htbp]
\renewcommand{\figurename}{\textbf{Figure}}
\renewcommand{\thefigure}{\textbf{\arabic{figure}}}
\centering\includegraphics[width=1\columnwidth]{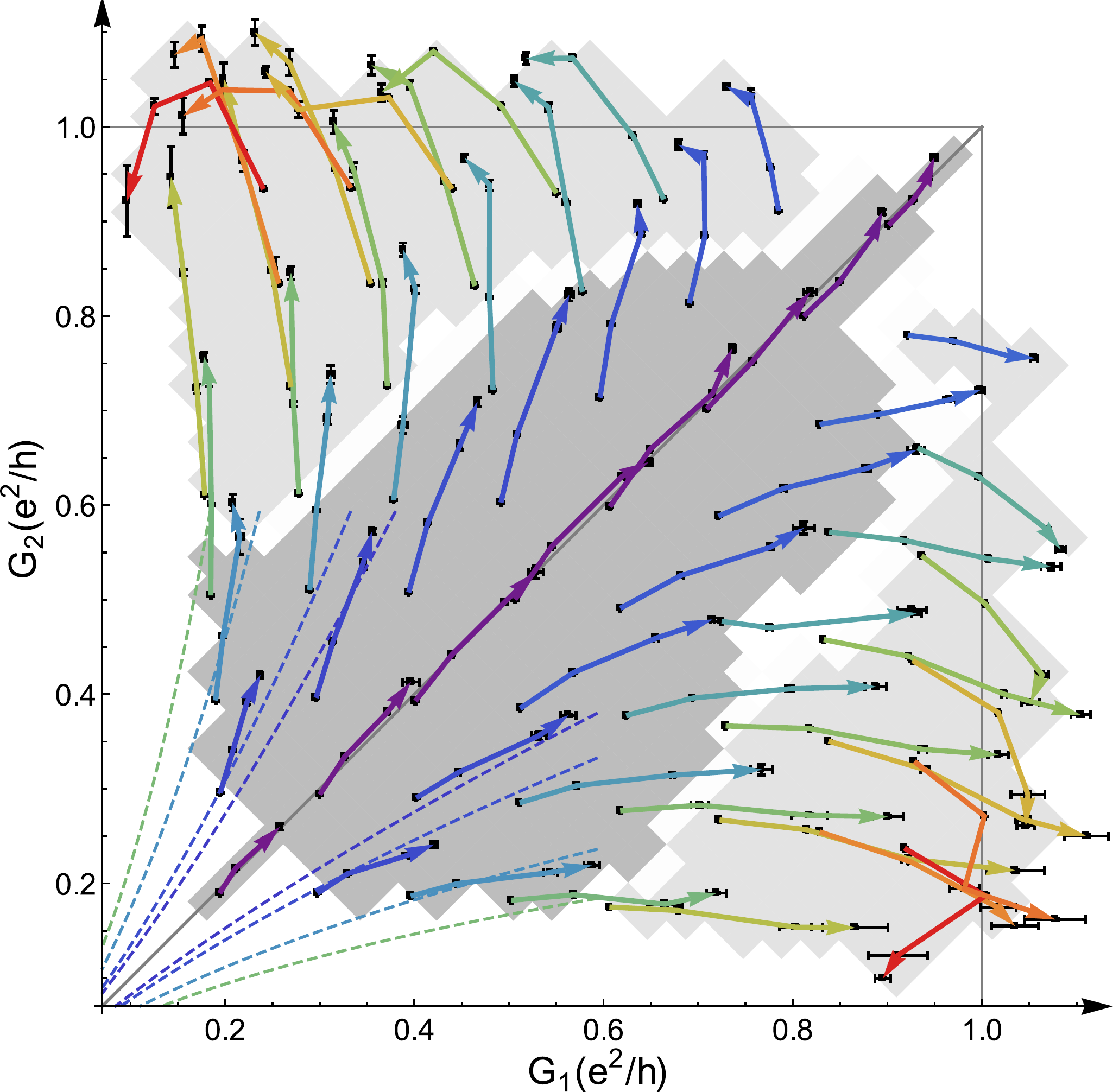}
\caption{\small
\textbf{Two-channel renormalization flow.} 
The `in-situ' (Kondo renormalized) conductances ($G_1,G_2$) measured at $T\simeq\{80,38,22,14\}$~mK at charge degeneracy are displayed as symbols, with a line connecting different temperatures of a same QPC$_{1,2}$ setting (characterized by the `intrinsic', unrenormalized, $\tau_{1,2}$), and a color code associated with $|\tau_1-\tau_2|$ (from purple for $|\tau_1-\tau_2|\simeq0$, to red for $|\tau_1-\tau_2|\simeq0.57$). 
The arrows pointing to the 14~mK conductance data points show the flow direction for decreasing temperatures. 
Indicative error bars are obtained by repeating the measurement at several nearby charge degeneracy points.
The 2CK (1CK) zone of influence is displayed as a gray (light gray) background.
The conductance flows predicted at small $G_{1,2}\ll e^2/h$, for the parameters corresponding to the crossed data line of the same color, are shown as dashed lines (Methods). 
\normalsize}
\label{fig-renormFlow}
\end{figure}

The 2CK phenomenology is directly revealed by the Kondo renormalization flow of the channels' coupling, as temperature is reduced. 
It is experimentally characterized by the (renormalized) `in-situ' conductances $G_{1,2}$.
We extract $G_1$ and $G_2$ separately by slightly opening QPC$_p$, with $G_p\ll G_{1,2}$ in order to minimize its effect (Methods). 
Figure~4 displays the $(G_1,G_2)$ renormalization flow for $T\simeq\{80,38,22,14\}$~mK. 
The continuous lines connect data points (symbols) obtained for identical $(\tau_1,\tau_2)$, with an arrow indicating the flow direction and a color corresponding to $|\tau_1-\tau_2|$ ($\tau_{1,2}\lesssim 0.12$ and $|\tau_1-\tau_2|\gtrsim 0.57$ are not included due to the small signal to noise).

First, note that asymmetric $G_{1,2}$ flow away from the symmetric line, exposing plainly the development of the predicted quantum phase transition across the symmetric quantum critical point \cite{Nozieres1980,Cox1998,Furusaki1995b,Pustilnik2004}.
Second, the renormalization flow also displays the predicted crossover from 2CK to 1CK behavior. 
The 2CK zone of influence, shown as a gray background in Fig.~4, is characterized by an increase of both $G_1$ and $G_2$ as $T$ is reduced. 
This occurs for $G_{1,2}\lesssim 0.5e^2/h$ (that is, $T\lesssim T_K$) or for relatively symmetric $G_{1,2}$.
The 1CK zone of influence, shown as a light gray background in Fig.~4, is characterized by the reduction of the smallest `in-situ' conductance as $T$ is lowered, while the largest further increases until reaching $\sim e^2/h$.
This occurs for asymmetric $G_{1,2}$ and only if the largest `in-situ' conductance is above $\sim 0.5e^2/h$, corresponding to an important screening of the pseudospin.
Note that the limit of one perfectly ballistic QPC was previously investigated in the context of dynamical Coulomb blockade \cite{Jezouin2013}.

Further information are disclosed by the experimental renormalization flow, including the temperature evolution of channel asymmetry.
Intriguingly, we also observe (Fig.~4) that the `in-situ' conductance of the most strongly coupled QPC can slightly overstep the standard quantum limit $e^2/h$. 
This overshoot is robust to experimental conditions, above noise level and not a simple calibration artifact (Methods). 

The present observation of the two-channel `charge' Kondo effect demonstrates that Kondo physics applies to the degenerate macroscopic quantum states of electrical circuits.
Our hybrid device allows full control and characterization of the Kondo parameters, and gives access to $(N\geq2)$-channel Kondo physics.
The implementation in the quantum Hall regime also opens the path to exploring the Kondo physics with anyonic quasiparticles, at fractional filling factors.
One limitation is the smallness of the charging energy $E_C$.
However, we anticipate that much higher $E_C$ are feasible by replacing the buried 2D electron gas with a surface conductor, such as graphene.

\bibliographystyle{nature}

\vspace{\baselineskip}
\small
{\noindent\textbf{Acknowledgments.}}
This work was supported by the ERC (ERC-2010-StG-20091028, \#259033) and the French RENATECH network.
We gratefully acknowledge E.~Boulat, J.~von Delft, S.~De Franceschi, L.~Glazman, D.~Goldhaber-Gordon, K.~Le Hur, A.~Keller, K.~Matveev, L.~Peeters, P.~Simon and G.~Zar\'and for the critical reading of our manuscript and discussions. 

{\noindent\textbf{Author Contributions.}} 
Z.I. and F.P. performed the experiment. 
Z.I., A.A. and F.P. analyzed the data. 
F.D.P. fabricated the sample. 
U.G. and A.C. grew the 2DEG. 
S.J. contributed to a preliminary experiment. 
F.P. led the project and wrote the manuscript with inputs from Z.I., A.A. and U.G.

{\noindent\textbf{Author Information.}}
Reprints and permissions information is available at www.nature.com/reprints.
The authors declare no competing financial interests.
Correspondence and requests for materials should be addressed to F.P. (frederic.pierre@lpn.cnrs.fr).
\normalsize 

\newpage
{\Large\noindent\textbf{METHODS}}

\small
{\noindent\textbf{Experimental setup.}} 
The measurements were performed using standard lock-in techniques, at frequencies below 100~Hz, in a dilution refrigerator.
Multiple filters along the electrical lines and two shields at the mixing chamber protect the sample from spurious high energy photons.

{\noindent\textbf{Sample.}} 
The sample is nanostructured by standard e-beam lithography in a 70~nm deep GaAs/Ga(Al)As two-dimensional electron gas of density $2.5 \times10^{11}~\mathrm{cm}^{-2}$ and mobility $10^6~\mathrm{cm}^2\mathrm{V}^{-1}\mathrm{s}^{-1}$. 
The metallic island is constituted of nickel (30~nm), germanium (60~nm) and gold (120~nm). 

{\noindent\textbf{Electronic temperature.}} 
The electronic temperature $T$ and the associated error bars are obtained from standard quantum shot noise measurements across both QPC$_{1,2}$ and, at $T \geq 38$~mK, also from the readings of a RuO$_2$ thermometer. 
The temperature stability is ascertained by measuring the electronic temperature before and after data acquisition, as well as with continuous RuO$_2$ readings.
For details on the noise measurement setup see the supplementary materials of ref.~\citenum{Jezouin2013b}.

{\noindent\textbf{Electronic level spacing in the metallic island.}} 
The typical energy spacing between electronic levels in the central metallic island is evaluated from the standard expression $\delta=1/(\nu_F \Omega)$, with $\Omega$ the island's volume and $\nu_F$ the electronic density of states per unit volume and energy in the metallic island.
Injecting the island's volume $\Omega \simeq 3~\mu\mathrm{m}^3$ and a typical density of states for metals $\nu_F\approx 10^{47}~\mathrm{J}^{-1}\mathrm{m}^{-3}$ (in gold, the main constituent, $\nu_F \simeq 1.14 \times 10^{47}~\mathrm{J}^{-1}\mathrm{m}^{-3}$), we find $\delta\approx k_B \times 0.2~\mu \mathrm{K}\lll k_B T$.
The very small electronic level spacing, more than four orders of magnitude smaller than the thermal energy $k_B T$, verifies the essential hypothesis $\delta\ll k_BT$ in the theory\cite{Matveev1991,Furusaki1995b,Matveev1995}.
To further demonstrate that, in general, the electronic level spacing in the island is fully negligible, one can compare $\delta$ with the electronic level energy width $h/\tau_\phi$, where $\tau_\phi$ is the electronic quantum coherence time.
Indeed, $\delta\ll h/\tau_\phi$ corresponds to a continuous electronic density of states.
The typical electron quantum coherence time is in the 10~ns range at low temperatures in similar diffusive metals\cite{Pierre2003} (see e.g. ref.~\citenum{Wellstood1994} for the measurement of $\tau_\phi$ in gold).
The corresponding electronic level energy width $h/\tau_\phi\sim k_B\times5~\mathrm{mK}\ggg\delta$ is therefore greater than the typical level spacing by approximately four orders of magnitude.

{\noindent\textbf{Interface metallic island - 2D electron gas.}} 
It is crucial to achieve a nearly perfect transmission of the outer electronic channel propagating along the edge of the buried 2DEG toward the central metallic island.
Here, we detail the procedure to precisely determine this transmission probability.
The notations are recapitulated in Extended Data Figure~1.
In the following, the lateral gates are fully depleted.
First, QPC$_{1,2,p}$ are set to the middle of the very flat and large ($\sim 0.4$~V) intermediate plateau at $\tau_{1,2,p}=1$ (thanks to the robust quantum Hall effect, see Extended Data Fig.~2c for the corresponding plateau across a lateral characterization gate) and we measure the corresponding $V_{ii}^{\tau_{1,2,p}=1}$ ($i\in \{1,2,p\}$). 
The transmission probability $\tau_{\Omega-i}$ of the outer edge channel from QPC$_{i}$ into the metallic island is then given by the expression
\begin{equation}
V_{ii}^{\tau_{1,2,p}=1} =(2-\tau_{\Omega-i})V_i/2+\tau_{\Omega-i}\frac{\tau_{\Omega-i}V_i/2}{\tau_{\Omega-1}+\tau_{\Omega-2}+\tau_{\Omega-p}}.\nonumber
\end{equation}
Note that we made absolutely sure there are no other ways than through the metallic island to go from QPC$_i$ to QPC$_j$, with $i\neq j$. This is done by etching trenches in the 2DEG underneath the island (see Fig.~1a and Extended Data Fig.~1).
Second, we eliminate calibration uncertainties by measuring the reflected signals $V_{ii}^{\tau_{1,2,p}=0}=V_i$ with QPC$_{1,2,p}$ disconnected (depleted).
The ratios $V_{ii}^{\tau_{1,2,p}=1}/V_{ii}^{\tau_{1,2,p}=0}$ give $\tau_{\Omega-i}$ independently of the injection and measurement chains calibrations:
\begin{equation}
\frac{V_{ii}^{\tau_{1,2,p}=1}}{V_{ii}^{\tau_{1,2,p}=0}} =(2-\tau_{\Omega-i})/2+\tau_{\Omega-i}\frac{\tau_{\Omega-i}/2}{\tau_{\Omega-1}+\tau_{\Omega-2}+\tau_{\Omega-p}}.
\end{equation}
With this approach, we obtain  $|1-\tau_{\Omega-i}|\lesssim 3 \times 10^{-4}$:
\begin{equation}
\tau_{\Omega-1} = 0.9997,~\tau_{\Omega-2} = 1.0003,~\tau_{\Omega-p} = 1.0001.\nonumber
\end{equation}
The outer edge channel is perfectly transmitted into the metallic island at our experimental accuracy.

{\noindent\textbf{Calibration of injection and measurement chains.}} 
In the same spirit as above, and now assuming $\tau_{\Omega-i}=1$, we normalize the signal $V_{ij}$ (see notations in Extended Data Fig.~1) by the signal $V_{ij}^{\tau_{i,j(k)}=1(0)}$ measured when setting $\tau_{i,j}=1$ with the other QPC disconnected, $\tau_{k}=0$.
For $i\neq j$, this gives
\begin{equation}
v_{ij} \equiv V_{ij}/V_{ij}^{\tau_{i,j(k)}=1(0)}=\frac{G_iG_j 2h/e^2}{G_1+G_2+G_p}.
\label{eqvij}
\end{equation}
The same information can also be extracted by solving the set of three equations for the reflected signals ($i=j$)
\begin{equation}
v_{ii} \equiv V_{ii}/V_{ii}^{\tau_{1,2,p}=0}=(1-G_ih/2e^2)+\frac{G_i^2 h/2e^2}{G_1+G_2+G_p}.
\label{eqvii}
\end{equation}
Note that if $G_p=0$, the measurements of $v_{11}$, $v_{22}$, $v_{12}$ and $v_{21}$ are redundant and only give access to $G_\mathrm{SET}=1/(G_1^{-1}+G_2^{-1})$, but not to $G_1$ and $G_2$ separately. 

{\noindent\textbf{Quantum point contacts characterization.}} 
Extended Data Fig.~2a(b) displays as a continuous line the measured `intrinsic' (not renormalized by Kondo effect or Coulomb blockade) transmission probability $\tau_{1(2)}$ of QPC$_{1(2)}$ versus the gate voltage $V_\mathrm{qpc1(2)}$ applied to one side of the corresponding split gate ($T\simeq 11.5$~mK, no dc bias voltage). 
The symbols indicate the QPC set points used in the experiment. 
Note that for larger (less negative) values of $V_\mathrm{qpc1,2}$, the `intrinsic' quantum point contact conductances exhibit a wide ($\sim 0.4$V) plateau, precisely at $e^2/h$ and robust to dc voltages within the explored range $|V_\mathrm{dc}|<100~\mu$V. This is followed by a second step up to $2e^2/h$ corresponding to the opening of a second  electronic (inner edge) channel (not shown but similar to the lateral characterization gate, see Extended Data Fig.~2c).
The insets in Extended Data Fig.~2a,b show the relative variation of the corresponding `intrinsic' QPC differential conductance with the applied dc bias voltage, up to $|V_\mathrm{dc}|=50~\mu\mathrm{V}\simeq 2 E_C/e$ and for $\tau_{1,2}\simeq \{0.06,0.47,0.93\}$ (data shifted vertically by $0.1$ for clarity).
The relatively small impact of dc bias voltage corroborates a point-like description of the quantum point contacts within the pertinent energy range, below $E_C$.
Note that the broad dip visible in the transmission across QPC$_1$ at larger split gate voltages $V_\mathrm{qpc1}$ (Extended Data Fig.~2a) has no impact at the used experimental set points.
In particular, it does not result in strongly energy dependent transmission probabilities (inset of Extended Data Fig.~2a and Extended Data Fig.~2e) and it has no impact on the dynamical Coulomb blockade low bias conductance suppression (Extended Data Fig.~2d). 
To perform the measurements in Extended Data Fig.~2a,b, both lateral characterization gates (Fig.~1a, colorized yellow for QPC$_2$, not colorized for QPC$_1$) were set to zero gate voltage. 
As shown Extended Data Fig.~2c, this corresponds to fully transmitting the two electronic edge channels across the lateral gates, thereby effectively short-circuiting the central metallic island (in normal operations the lateral characterization gates are set to $\approx -0.4$~V in order to deplete the 2DEG underneath, for further details regarding the lateral characterization gates `switch' operation see the supplementary information in ref.~\citenum{Parmentier2011}).
Extended Data Fig.~2d shows as continuous lines the differential conductance across QPC$_{1,2}$ measured at 22~mK as a function of dc voltage with the nearby lateral characterization gate set to deplete the 2DEG (biased at $\approx -0.4$~V, as when exploring the 2CK physics) while the lateral gate on the opposite side of the metallic island is set to transmit the two edge channels (biased at $\approx 0$~V), as illustrated schematically.
The central conductance dip at low dc voltage corresponds to the dynamical Coulomb blockade suppression of the conductance\cite{Jezouin2013,Parmentier2011}, while the flat plateaus at large dc voltages are used to extract the `intrinsic' transmission probabilities $\tau_{1,2}$ here displayed as horizontal dashed lines.
The precise values of the `intrinsic' transmission probabilities $\tau_{1,2}$ at the experimental set points, and their relative increase $\Delta \tau_{1,2}/\tau_{1,2}$ between zero bias and $\pm 50~\mu\mathrm{V}$ (corresponding to our estimated experimental uncertainty on $\tau$), are recapitulated in Extended Data Fig.~2e.

{\noindent\textbf{Capacitive cross-talk.}} 
Changing the gate voltage controlling one QPC also slightly affects the other ones. 
This cross-talk is determined precisely using the lateral characterization gates, from the shift in gate voltage of the QPC `intrinsic' conductance curves shown Extended Data Fig.~2a,b.
Thanks to the relatively important distances (several micrometers) between QPCs (compared to small quantum dots) the cross-talk correction is small, typically a few percent.
We take into account the small capacitive cross-talk correction during data acquisition.

{\noindent\textbf{Charging energy characterization.}} The charging energy $E_C=e^2/2C\simeq k_B\times290$~mK is obtained from the measured Coulomb diamonds displayed Extended Data Fig.~3.

{\noindent\textbf{Conductance peak reproducibility.}} Although a single period of $G_\mathrm{SET}(V_g)$ is shown Fig.~1c, we systematically measured several nearby periods for each configuration.
Extended Data Fig.~4 displays as symbols several consecutive periods measured at base temperature $T=11.5$~mK for the same configuration $\tau=0.93$ shown in Fig.~1c, together with the quantitative theoretical prediction of Eq.~\ref{GsetFM} (continuous line).
In practice we take the average of the maximum conductance (at charge degeneracy) measured for different periods, and we estimate the experimental uncertainty (s.e.m.) from the the scatter between values.
For some relatively rare combinations of QPC settings, temperatures and precise gate voltages, we find anomalously small values of the maximum conductance with respect to the overall experimental standard deviation.
We systematically eliminate such anomalous data points, which can often be attributed to charge jumps in the sample vicinity, by considering only the data within a window of four times the overall experimental standard deviation.
This automatic procedure removes approximately 10\% of the measured local conductance maximums.
In Figs.~2,3, the extracted uncertainty (not shown) is smaller than the symbols.  
In Fig.~4, the extracted experimental uncertainty is displayed as error bars.

{\noindent\textbf{Theoretical expression of $\mathbf G_\mathrm{SET}$ and $\mathbf G_{1,2}$ at $\mathbf\tau_{1,2}\ll 1$ ($T\gg T_K$).}} 
In the limit $T\gg T_K$ (also corresponding to the tunnel regime $\tau_{1,2}\ll 1$) the two Kondo channels are independent from one another since they only weakly screen the pseudospin-$1/2$.
Consequently the `in-situ' (Kondo renormalized) conductances $G_{1,2}\ll e^2/h$ renormalize independently, increasing as temperature is reduced near charge degeneracy ($\delta V_g\approx 0$) due to the Kondo effect.
Theory predicts that the standard expression Eq.~\ref{eqseq} for independent sequential tunneling events holds provided that the `intrinsic' transmission probabilities $\tau_{1,2}$ in $G_\infty=(e^2/h)/(\tau_1^{-1}+\tau_2^{-1})$ are substituted by the Kondo renormalized values\cite{Furusaki1995b}
\begin{equation}
\tau_{1,2} \rightarrow \pi^2/\log^2 (\max \{T,2 E_C |\delta V_g/\Delta|/k_B\}/\alpha T_{K1,2}),
\label{eqtau}
\end{equation}
where $\alpha$ is a numerical factor depending on the precise definition of $T_{K1,2}=T_K(\tau_{1,2})$ (Eq.~\ref{eqTKtunnel}, see `Rescaling procedure' below for the determination of $\alpha$).
Note that the substitution Eq.~\ref{eqtau} leaves the width of the conductance line shape essentially proportional to temperature, although slightly narrower in reasonable agreement with the data.
Consequently, at a good approximation in the tunnel regime, the Kondo effect essentially results in an increased value of the parameter $G_\infty$ in Eq.~\ref{eqseq}.
In this spirit, we have fitted the tunnel data shown Fig.~1b (symbols) using Eq.~\ref{eqseq} with the temperature and $G_\infty$ as free parameters (continuous line).
In Fig.~2b, at charge degeneracy ($\delta V_g=0$), the displayed theoretical (thy) $T>>T_K$ prediction (violet dashed line) is given by
\begin{equation}
G_{\mathrm{SET}}^{\mathrm{thy~}T/T_K\gg 1}\left(T/T_K\right)=9.62 \frac{e^2}{h} \log^{-2}\left(\frac{T}{0.0037 T_K}\right).
\label{eqGsetTgd}
\end{equation}
In Fig.~4, the displayed predictions for the renormalization flow at small $G_{1,2}\lesssim 0.6 e^2/h$ (dashed lines with the same color code as the corresponding data) are calculated without additional fit parameters, using $G_{1,2}=2 G_\mathrm{SET}^{\mathrm{thy~}T/T_K\gg 1}\left(T/T_{K1,2}\right)$, with $T_{K1,2}=T_K(\tau=\tau_{1,2})$ given by the previously extracted experimental scaling temperature shown in the inset of Fig.~2b, and with $G_\mathrm{SET}^{\mathrm{thy~}T/T_K\gg 1}$ given by Eq.~\ref{eqGsetTgd}.  

{\noindent\textbf{Theoretical expression of $\mathbf{G_\mathrm{SET}}$ at $\mathbf{\tau_{1,2}\approx 1}$.}} 
The quantitative expression of $G_\mathrm{SET}$ has been established for arbitrary offsets from the charge degeneracy point ($\delta V_g$), in the limit where both QPC$_{1,2}$ are set close to the ballistic limit ($1-\tau_{1,2}\ll 1$) and for low temperatures with respect to the charging energy $k_BT\ll E_C$ (ref.~\citenum{Furusaki1995b}, based on the theoretical framework developed in ref.~\citenum{Matveev1995}).
The prediction shown as a continuous line in Fig.~1c is obtained quantitatively, without fit parameters, from the following theoretical expression (Eqs.~38, 26 and A9 in ref.~\citenum{Furusaki1995b}):
\begin{equation}
G_\mathrm{SET}  =  \frac{e^2}{2 h} \Big[ 1- \frac{\pi^3 \gamma \Gamma_+ k_B T}{16 E_C}  -\int^{\infty}_0 \frac{\Gamma^2_-/\cosh^2(x)}{(x\pi^2k_BT/\gamma E_C)^2+\Gamma^2_-}\mathrm{d}x \Big], \label{GsetFM}
\end{equation}
with $\gamma \simeq \exp(0.5772)$ and
\begin{equation}
\Gamma_{\pm} = 2-\tau_1-\tau_2 \pm 2 \sqrt{(1-\tau_1)(1-\tau_2)} \cos(2 \pi \delta V_g/\Delta).\nonumber
\end{equation}
Note that we have supplemented Eq.~38 of ref.~\citenum{Furusaki1995b} with the small correction proportional to $k_BT/E_C$ in Eq.~A9, following the same procedure used in Fig.~2 of ref.~\citenum{Furusaki1995b}.
The function $\Gamma_{-}$ reduces to zero when the sample is set to display the 2CK effect ($\tau_1=\tau_2$ and $\delta V_g=0$).
Instead, the integral term with $\Gamma_{-}$ in Eq.~\ref{GsetFM} determines the crossover from quantum criticality, as further discussed in the next section.
In symmetric situations ($\tau_1=\tau_2$) and at the degeneracy point ($\delta V_g=0$), all the temperature dependence describing the flow toward the 2CK state (quantum critical point) results from the term proportional to $\Gamma_{+}$ in Eq.~\ref{GsetFM}.
Without additional hypothesis than $\tau\equiv\tau_1=\tau_2$ and $\delta V_g=0$, Eq.~\ref{GsetFM} can be reformulated as a universal scaling function, whose value tends linearly toward the quantum critical point $e^2/2h$ when the temperature goes to 0:
\begin{equation}
G_\mathrm{SET}^{T\ll T_K}(T/T_K^{1-\tau\ll 1},\delta V_g=0)  = \frac{e^2}{2 h} \left( 1- T/2T_K^{1-\tau\ll 1} \right),  \label{GsetTllTK}
\end{equation}
with the Kondo scaling temperature defined as
\begin{equation}
T_K^{1-\tau\ll 1}=\frac{2 E_C}{\pi^3 \gamma k_B (1-\tau)}. \label{TKtausim1}
\end{equation}
Note that although for small enough $1-\tau$ the Kondo temperature can become larger than the charging energy $E_C$, the latter remains a high energy cutoff for the Kondo physics since at larger energies (e.g. $k_BT\gtrsim E_C$) additional charge states of the island become accessible.
Note also that equally valid definitions of the Kondo temperature can differ by a constant multiplicative factor: replacing $T/2T_K^{1-\tau\ll 1}$ in Eq.~\ref{GsetTllTK} by $\alpha T/2T_K^{1-\tau\ll1}$ would change the expression of $T_K^{1-\tau\ll1}$ by the multiplicative factor $\alpha$.
Here, the definition of $T_K^{1-\tau\ll1}$ was chosen such that $G_\mathrm{SET}^{T\ll T_K}(T=T_K^{1-\tau\ll1},\delta V_g=0)=0.5\times e^2/2h$ (although $T=T_K^{1-\tau\ll1}$ is beyond the range of validity of Eq.~\ref{GsetTllTK}).
The red short-dashed line displayed in the main panel of Fig.~2b is the quantitative prediction for $G_\mathrm{SET}(T)$ calculated with Eq.~\ref{GsetTllTK} for $\tau=0.86$ and $E_C=k_B\times290~$mK.
It was rescaled in $T/T_K$ using the same experimental scaling temperature $T_K(\tau=0.86)\simeq1.4$~K as the $\tau=0.86$ data (and not $T_K^{1-\tau\ll1}(\tau=0.86)\simeq0.075$~K) to allow a direct comparison data/theory in Fig.~2b. 
In the inset of Fig.~2b, a constant multiplicative factor is applied to $T_K^{1-\tau\ll 1}\propto1/(1-\tau)$ (red short-dashed line) to match the experimental scaling temperature at $\tau\approx1$.

{\noindent\textbf{Predictions for the crossover from quantum criticality.}} 
In this section, we show that the predictions of Eq.~\ref{GsetFM} correspond to generic expectations for the crossover from quantum criticality\cite{Cox1998,Pustilnik2004,Sela2011}.
An asymmetry between Kondo channels ($\tau_1-\tau_2\neq0$) or a lifting of the charge pseudospin degeneracy ($\delta V_g\neq0$) is predicted to destroy the unstable 2CK state at vanishing temperatures; and a crossover from non-Fermi liquid (quantum critical) to Fermi liquid behavior is expected to take place as temperature is reduced. 
The corresponding crossover temperature is generically expected\cite{Cox1998,Pustilnik2004,Sela2011} to depend quadratically on the strength of the perturbations near the symmetric ($\tau_1=\tau_2$, $\delta V_g=0$) quantum critical point.
The theoretical prediction of Eq.~\ref{GsetFM} describes quantitatively the crossover from quantum criticality for the present two-channel `charge' Kondo effect, in the presence of an asymmetry between the two channels and/or of a pseudospin energy splitting.
As generically expected, Eq.~\ref{GsetFM} predicts that any perturbation ($\tau_1\neq\tau_2$ and/or $\delta V_g\neq0$) results in a SET conductance vanishing in the low temperature limit as $T^2$, the standard Fermi-liquid power law (see also Eq.~39 in ref.~\citenum{Furusaki1995b}).
The crossover behavior is described by a single function, independent of the perturbations (channel asymmetry $\delta\tau=\tau_1-\tau_2$, energy splitting $\Delta E=2E_C\delta V_g/\Delta$, or both simultaneously) that are encapsulated in the parameter $\Gamma_-$, thereby corroborating the universal behavior put forward in ref.~\citenum{Sela2011}.
The crossover temperature $T_\mathrm{co}$ can be extracted from Eq.~\ref{GsetFM}.
Here it is defined as the temperature at which $G_\mathrm{SET}=0.5\times e^2/2h$ (assuming a fully developed 2CK state in absence of perturbation, ie neglecting the term proportional to $\Gamma_+T/E_C$ in Eq.~\ref{GsetFM}).
At $\delta V_g=0$ and for small $\delta\tau\ll2-\tau_1-\tau_2$, one obtains from Eq.~\ref{GsetFM} the crossover temperature $T_\mathrm{co}(\Delta E=0,\delta\tau)\simeq(\gamma^2\pi/4) T_K^{1-\tau\ll 1}(\delta\tau)^2$, corresponding to generic predictions (detailed in e.g. ref.~\citenum{Sela2011}).
At $\delta\tau=0$ and for small $\delta V_g\ll\Delta$, one obtains from Eq.~\ref{GsetFM} the crossover temperature $T_\mathrm{co}(\Delta E,\delta\tau=0)\simeq(4/\pi^3) T_K^{1-\tau\ll 1}(\Delta E/k_B T_K^{1-\tau\ll 1})^2$, corresponding to generic predictions\cite{Sela2011}.

{\noindent\textbf{Rescaling procedure.}}
We show Fig.~2b that the data $G_\mathrm{SET}(T,\tau)$ for symmetric QPCs ($\tau\simeq \tau_1\simeq \tau_2$) can be rescaled into a single curve $G_\mathrm{SET}(T/T_K)$.
To illustrate the procedure let us consider two successive transmissions $\tau$ and $\tau'$ with a conductance overlap such that one can find two data points $G_\mathrm{SET}(T,\tau)= G_\mathrm{SET}(T',\tau')$. 
The existence of a universal scaling law directly implies $T_K(\tau')/T_K(\tau)=T'/T$. 
If such a law exists, then the rescaled data at $\tau$ and $\tau'$ should match on the full range of conductance overlap.
This scheme does not apply directly for the three lowest transmission probabilities $\tau \simeq \{0.06,0.125,0.245\}$, since there is no conductance overlap. However, theory predicts\cite{Furusaki1995b} in the corresponding limit $T\gg T_K$ that $G_\mathrm{SET}(T)\propto \log^{-2} (T/\alpha T_K)$. 
Using this expression to fit and extrapolate the $\tau <0.25$ data points (dashed line in Fig.~2b), we can apply the above procedure. 
Note that $T_K(\tau)$ is extracted only up to an overall prefactor. 
Following standard usage\cite{Goldhaber-Gordon1998b}, we set this prefactor such that $G_\mathrm{SET}(T/T_K=1)=0.5\times e^2/2h$ (half the 2CK state conductance).

{\noindent\textbf{Absence of numerical renormalization group calculations for $\mathbf{G_\mathrm{SET}(T/T_K)}$.}}
To the best of our knowledge, there are no available numerical calculations for the measured conductance $G_\mathrm{SET}$.
Consequently, there is no theoretical prediction to compare with the experimentally extracted scaling curve $G_\mathrm{SET}(T/T_K)$ shown Fig.~2b, beyond the limits of large or small $T/T_K$.
Quoting the authors of ref.~\citenum{Pustilnik2004}, the root of the difficulty ``is that there is no mapping between the conductance across the island ($G_\mathrm{SET}$) and the electron scattering cross-section in the generic two-channel Kondo model".
Hopefully, future numerical works, adapted to the present charge Kondo implementation, will fill this gap and allow a full quantitative comparison data-theory, including at intermediate values of $T/T_K$.

{\noindent\textbf{Extracting separately the `in-situ' conductances $\mathbf{G_1}$ and $\mathbf{G_2}$ with QPC$\mathbf{_p}$.}}
The SET conductance, with QPC$_p$ disconnected, only gives access to the series combination of the `in-situ' (Kondo renormalized) conductances $G_1$ and $G_2$ ($G_\mathrm{SET}=1/(G_1^{-1}+G_2^{-1})$).
This is sufficient in symmetric configurations $G_1\simeq G_2$, but not to extract the full renormalization flow shown in Fig.~4.
For this purpose we use an additional probe QPC$_p$.
To minimize the effect of this probe, it is set to a relatively small coupling with respect to $G_1$ and $G_2$.
In practice, $1/150 < G_p/\mathrm{min}(G_1,G_2) < 1/6$, with the largest values corresponding to the most asymmetric configurations between QPC$_{1,2}$.
As easily checked from Eq.~\ref{eqvij}, this gives access directly to $G_1/G_2= v_{p1}/v_{p2}$ (or equivalently, $G_1/G_2= v_{1p}/v_{2p}$).
Solving Eqs.~\ref{eqvij} and \ref{eqvii}, with the measured $v_{ij}$ gives all three `in-situ' conductances $G_{1,2,p}$ (provided that $G_{1,2,p}\neq 0$).

{\noindent\textbf{`In-situ' conductances above the standard quantum limit $\mathbf{e^2/h}$.}}
Some of the `in-situ' conductances displayed Fig.~4 slightly overstep the standard quantum limit $e^2/h$ for asymmetric QPCs configurations and at low temperatures.
Although the standard quantum limit applies to a single quantum channel connected to voltage biased reservoirs (in contrast, the central metallic island is floating) and in the absence of interactions, to the best of our knowledge such a striking behavior was never observed.
In principle, a partial transmission of the second (inner) edge channel across the QPC could provide a simple explanation for the observation of an in-situ conductance above $e^2/h$.
However this is unlikely since the second electronic channel is initially completely reflected, separated from the full opening of the first (outer) channel by a plateau very wide and very robust to dc voltage (tested up to $|V_\mathrm{dc}|\simeq 100~\mu\mathrm{V}\approx 4E_C/e$). 
Note that we checked in-situ that the lateral characterization gates set to reflect the two edge channels ($\tau_\mathrm{lcg}=0$, at $V_\mathrm{lcg}\approx-0.4$~V), as well as QPC$_p$ when initially disconnected, remain in this configuration in presence of the charge Kondo effect.
It is also noteworthy that in the present experimental configuration, in the integer quantum Hall regime at filling factor $\nu=2$, the current between the metallic island and the QPCs is carried by two copropagating quantum Hall channels that are coupled by the Coulomb interaction.
However, for the short distance island-QPC and very low temperatures in the present experimental investigation, this coupling is expected to be negligible\cite{leSueur2010}. 
A similar transient overshoot is predicted in the related Luttinger liquid problem (at $K<1/2$, see Fig.~1 in ref.~\citenum{Fendley1995}), which corresponds to an `in-situ' single channel differential conductance above $e^2/h$ (in the context of the Luttinger liquid-dynamical Coulomb blockade mapping \cite{Safi2004,Jezouin2013}).

We here show that our intriguing observation is well above the noise level, that the same result is obtained with different sets of measurements, and that it is robust with respect to injection voltage and to the coupling of QPC$_p$.
For this purpose we focus on the set point $(\tau_1=0.76,\tau_2=0.93)$ at $T\simeq 14$~mK.
Extended Data Fig.~5a,b,c,d show the normalized transmitted signals $v_{ij}$ with $i\neq j$ and the reflected signal $2-v_{pp}$, measured in the linear response regime with here $V_i\simeq 1.15~\mu\mathrm{V_{rms}}<k_BT/e$.
The displayed statistical error bars are here obtained by repeating the measurements ten times in a row for each data point.
Possible charge offsets jumps are ruled out from the reproducibility of the two displayed consecutive sweeps ($V_g$ increasing and decreasing).
Note that the same normalized data are found for the reciprocal signals $v_{ij}\simeq v_{ji}$.
Note also that QPC$_p$ is here set to a different tuning (with a higher conductance) than the corresponding data point in Fig.~4.
First, we extract $G_{1,2,p}$ solving Eq.~\ref{eqvij} with only the transmitted signals ($v_{ij}$ with $j\neq i$). 
Averaging all the data ($V_g$ increasing and decreasing, and reciprocal signals) at the degeneracy point, we find
\begin{eqnarray}
G_1 &=& 0.508\pm 0.003~e^2/h,\nonumber\\
G_2 &=& 1.11 \pm 0.02~e^2/h,\nonumber\\
G_p &=&0.0387 \pm 0.0005~e^2/h.\nonumber
\end{eqnarray}
Second, we show that the same result is obtained with a different set of measurements, now involving also Eq.~\ref{eqvii}. 
We use $v_{12}/2$ and $2(1-v_{pp})$ (Extended Data Fig.~5d) corresponding to $1/(G_1^{-1}+G_2^{-1})$ and $G_p$, respectively, in the limit $G_p\ll G_{1,2}$, as well as $v_{1p}/v_{2p}=G_1/G_2$ Extended Data Fig.~5e).
This gives the consistent values
\begin{eqnarray}
G_1 &=& 0.510\pm 0.002~e^2/h,\nonumber\\
G_2 &=& 1.107 \pm 0.006~e^2/h,\nonumber\\
G_p &=&0.0395 \pm 0.0002~e^2/h.\nonumber
\end{eqnarray}
In fact, no averaging is required to show that $G_2>e^2/h$ well beyond the noise level. 
Focusing on the smallest signal $v_{1p}/v_{2p}=G_1/G_2$, we observe directly in Extended Data Fig.~5e that every data point near $\delta V_g \approx 0$ is below the red line displaying the ratio for which $G_2=e^2/h$.
Every data point therefore corresponds to $G_2>e^2/h$.
To further confirm $G_2>e^2/h$, we have also checked that this observation is robust with respect to injection voltage $V_i$ and to the value of $G_p$, as shown Extended Data Fig.~6a,b.

\begin{figure*}[p]
\renewcommand{\figurename}{\textbf{Extended Data Figure}}
\renewcommand{\thefigure}{\textbf{1}}
\centering\includegraphics [width=1\textwidth]{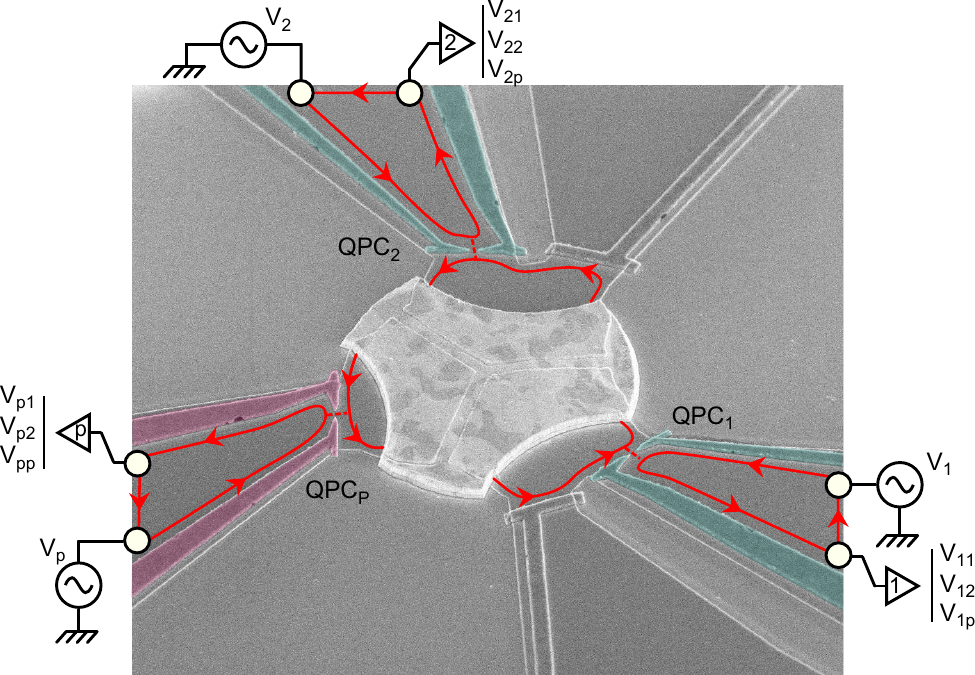}
\caption{\small
\textbf{Measurement schematic.} 
Schematic of the measurement setup showing explicitly the nine different and simultaneously measured signals. 
$V_{ij}$ ($i,j\in\{1,2,p\}$) is the voltage measured with amplification chain $i$ in response to the injected voltage $V_j$.
Trenches etched in the 2DEG in the form of a Y can be seen through the metallic island.
\normalsize}
\end{figure*}

\begin{figure*}[p]
\renewcommand{\figurename}{\textbf{Extended Data Figure}}
\renewcommand{\thefigure}{\textbf{2}}
\centering\includegraphics [width=1\textwidth]{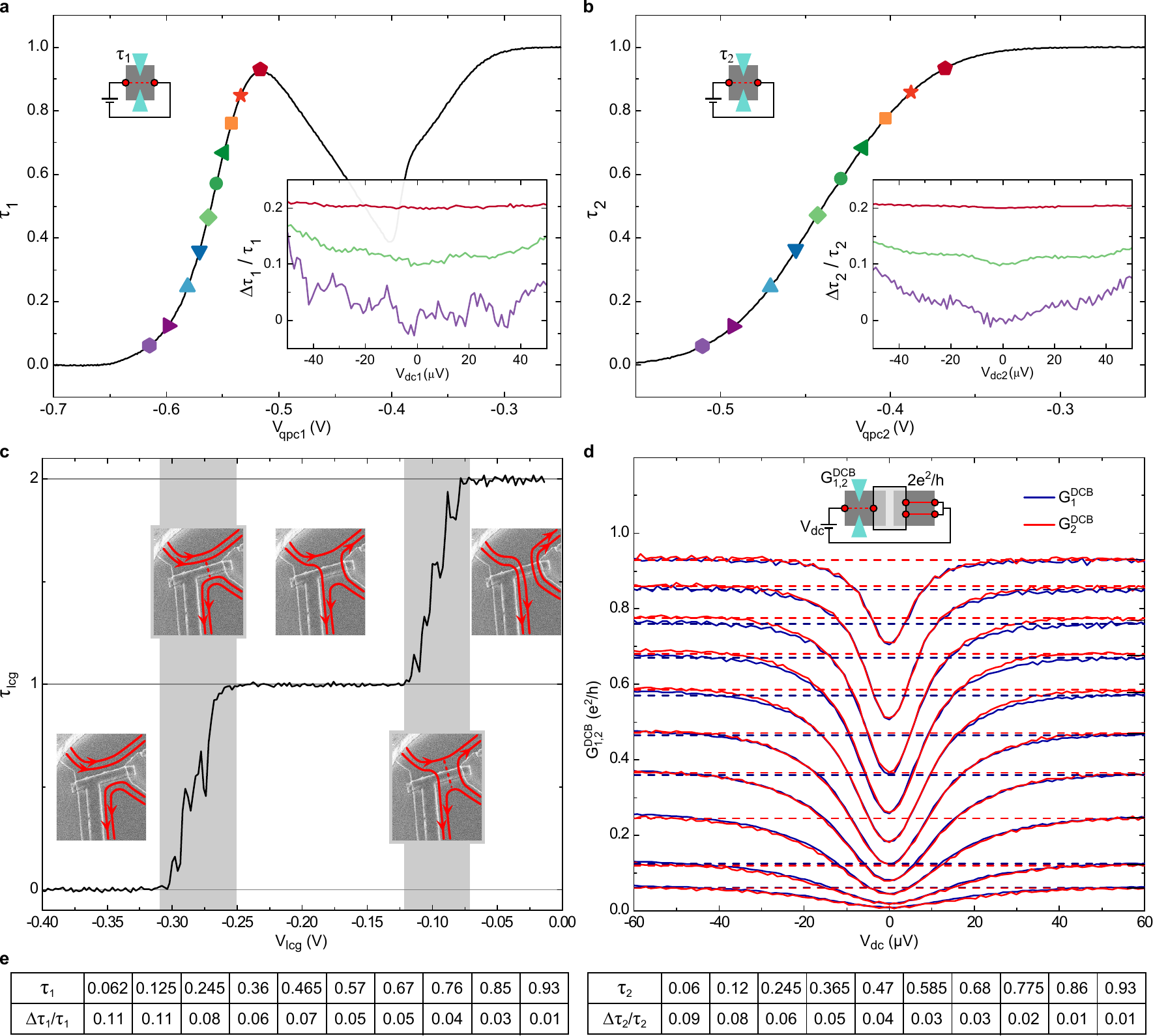}
\caption{\small
\textbf{Quantum point contact characterization.} 
\textbf{a},\textbf{b}, `Intrinsic' transmission probability across QPC$_1$ ($\tau_1$; a) and QPC$_2$ ($\tau_2$; b) measured at $11.5$~mK (in the linear regime, without dc bias) by opening the QPC lateral characterization gate (see equivalent schematic in top left insets), and plotted versus the voltage applied to the split gate tuning the QPC. 
The experimental transmission set points in the main text are indicated by symbols. 
Inset, relative variation of the transmission probability with dc bias voltage, shifted vertically for clarity, for $\tau_{1,2}\simeq \{0.06,0.47,0.93\}$ from bottom to top respectively. 
The larger noise in the inset of panel \textbf{a} (mostly visible for $\tau_1\simeq 0.06$) is from the amplification chain.
\textbf{c}, `Intrinsic' conductance across one lateral characterization gate in units of $e^2/h$ ($\tau_\mathrm{lcg}$, here adjacent to QPC$_1$) plotted versus lateral gate voltage $V_\mathrm{lcg}$. Increasing $V_\mathrm{lcg}$ results in the successive full opening of two electronic channels, as schematically illustrated. 
In practice, we close (open) the lateral characterization gates, corresponding to $\tau_\mathrm{lcg}=0$ ($\tau_\mathrm{lcg}=2$), by applying $V_\mathrm{lcg}\approx-0.4$~V ($V_\mathrm{lcg}=0$~V).  
Grey shaded areas correspond to the partial opening of one of the channel, a configuration not used in the experiment.
\textbf{d}, Conductance of the QPCs measured at $T=22$~mK versus dc voltage (continuous lines) with the adjacent lateral gate closed ($\tau_\mathrm{lcg}=0$) and the lateral characterization gate opposite to the metallic island set to full transmission ($\tau_\mathrm{lcg}=2$), which corresponds to the displayed schematic circuit. 
The low bias dips result from conductance suppression by the dynamical Coulomb blockade, while the high bias plateaus correspond to the `intrinsic' transmission probabilities $\tau_{1,2}$ (horizontal dashed lines).
\textbf{e}, `Intrinsic' transmission probabilities $\tau_{1,2}$ at the experimental set points used in the main text, together with their relative increase $\Delta \tau_{1,2}/\tau_{1,2}$ between $V_\mathrm{dc1,2}=0$ and $|V_\mathrm{dc1,2}|=50~\mu$V. 
This increase is the main experimental factor of uncertainty on the determination of $\tau_{1,2}$. 
\normalsize}
\end{figure*}

\begin{figure*}[p]
\renewcommand{\figurename}{\textbf{Extended Data Figure}}
\renewcommand{\thefigure}{\textbf{3}}
\centering\includegraphics [width=0.6\textwidth]{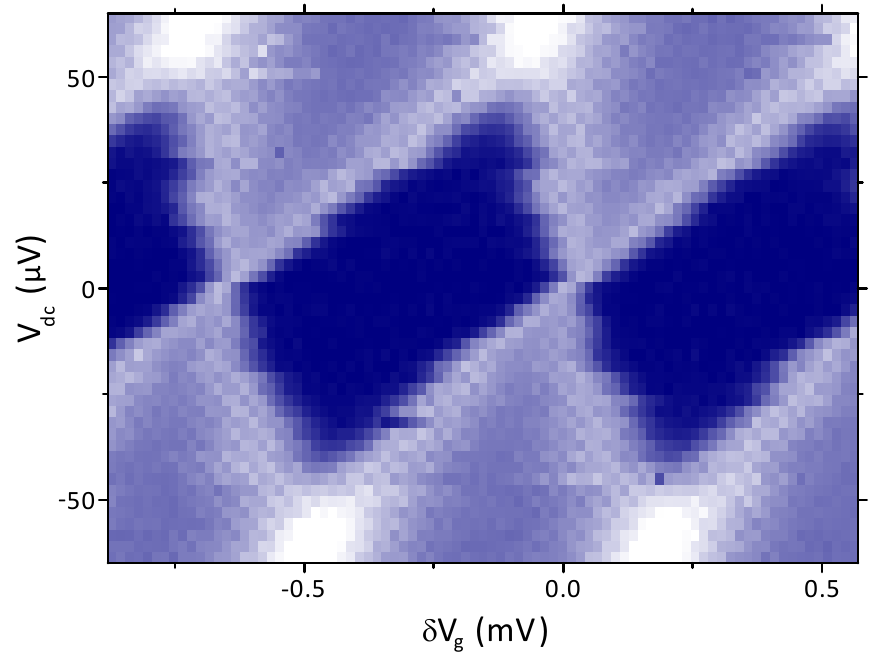}
\caption{\small
\textbf{Coulomb diamonds.} 
The conductance $G_\mathrm{SET}$ (brighter for larger $G_\mathrm{SET}$) is displayed versus gate and dc voltages ($\delta V_g$ and $V_\mathrm{dc}$, respectively), with both QPCs set to a low transmission probability.
The Coulomb diamonds (darker) correspond to a charging energy $E_C=e^2/2C\simeq k_B\times 290~$mK.
\normalsize}
\end{figure*}

\begin{figure*}[p]
\renewcommand{\figurename}{\textbf{Extended Data Figure}}
\renewcommand{\thefigure}{\textbf{4}}
\centering\includegraphics [width=0.75\textwidth]{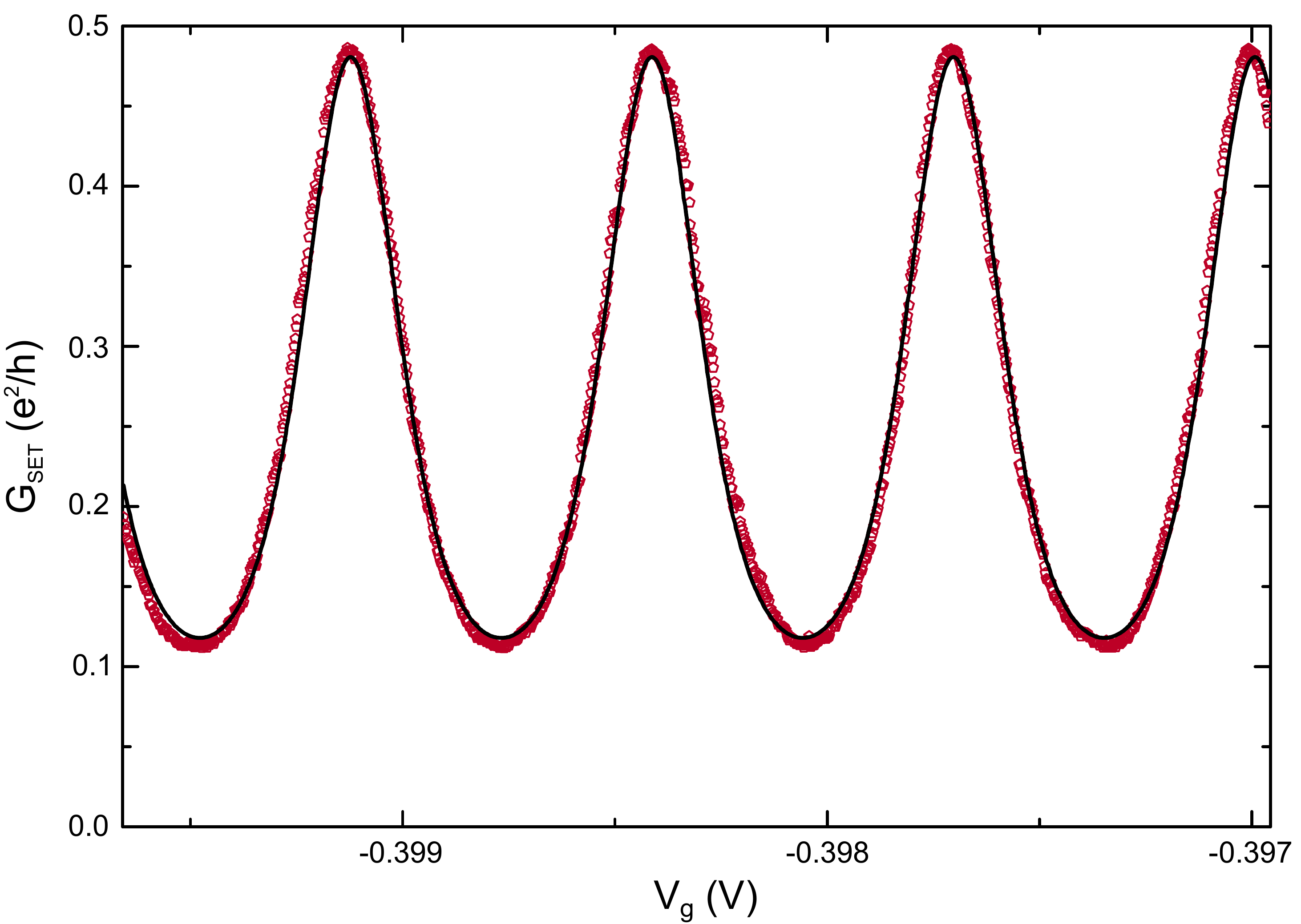}
\caption{\small
\textbf{Reproductibility of conductance oscillations.} 
Conductance across the hybrid SET when sweeping the gate voltage across several periods for the symmetric configuration $\tau_1\simeq\tau_2\simeq0.93$ and at base temperature $T\simeq11.5$~mK (one period shown in Fig.~1c). 
The symbols display the measurements, the continuous line is the quantitative prediction of Eq.~\ref{GsetFM}.
\normalsize}
\end{figure*}

\begin{figure*}[p]
\renewcommand{\figurename}{\textbf{Extended Data Figure}}
\renewcommand{\thefigure}{\textbf{5}}
\centering\includegraphics [width=1\textwidth]{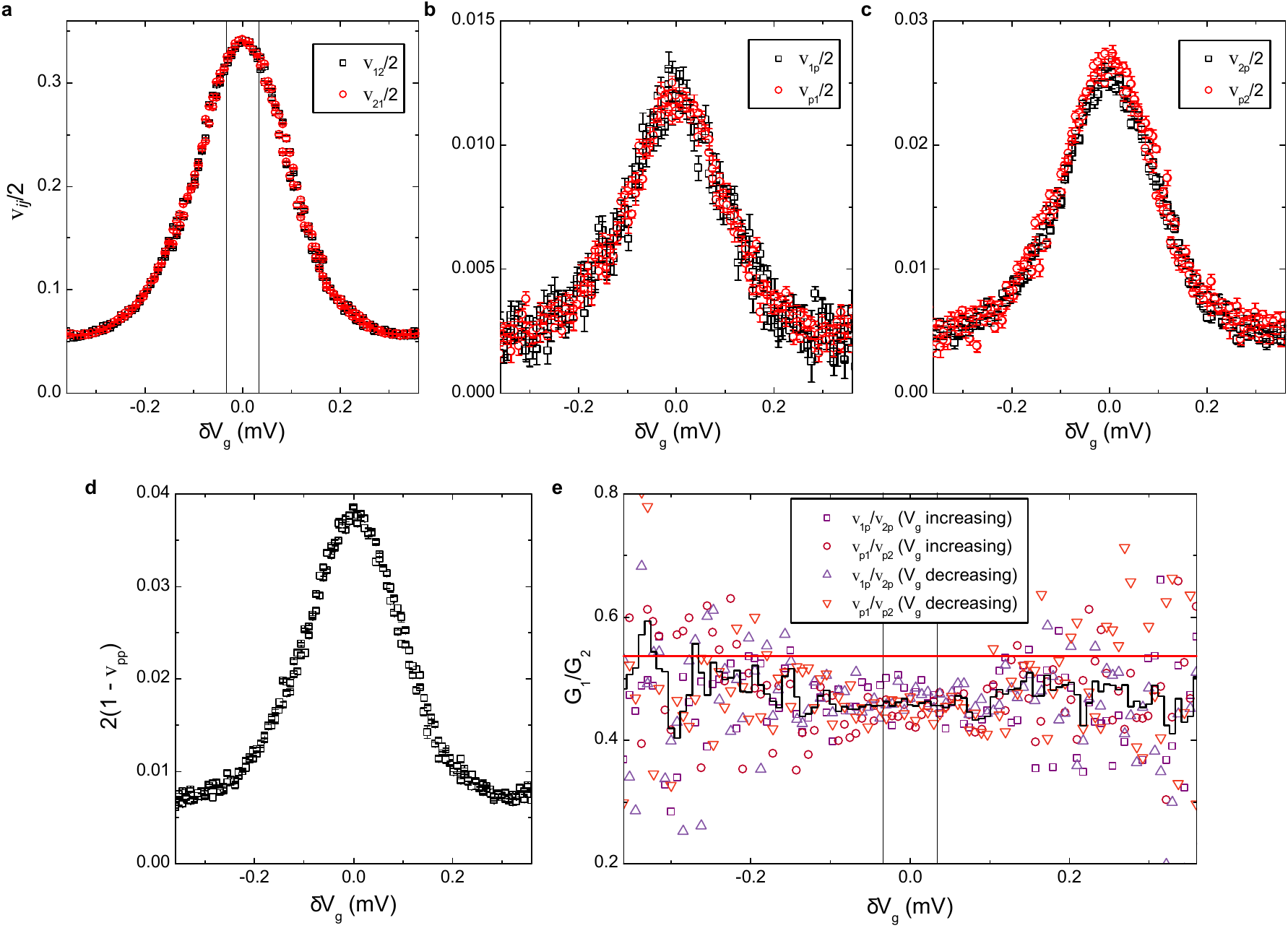}
\caption{\small
\textbf{Observation of `in-situ' conductances overstepping the standard quantum limit $\mathbf e^2/h$.} 
The displayed Coulomb peaks were measured at $T\simeq 14$~mK for the asymmetric QPCs configuration $(\tau_1 \simeq 0.77,\tau_2 \simeq 0.93)$. 
Two sweeps ($V_g$ increasing and decreasing) are shown for each measurement.
\textbf{a,b,c}, Symbols are the normalized transmitted signal $v_{i,j}/2$, with $i\neq j$ (Eq.~\ref{eqvij}) versus gate voltage. 
Each panel displays the two reciprocal signals $v_{i,j}$ and $v_{j,i}$. 
The vertical lines in panel \textbf{a} are visual markers used in panel \textbf{e}.
\textbf{d}, Symbols are the normalized reflected signal at the probe QPC$_p$ ($2(1-v_{pp})$, corresponding to $G_p h/e^2$ in the limit $G_p\ll G_{1,2}$).
\textbf{e}, Symbols are the in-situ conductances ratio $G_1 / G_2$, measured from both $v_{1p}/v_{2p}$ and $v_{p1}/v_{p2}$. 
For each measurement, the two sweeps ($V_g$ increasing and decreasing) are shown with different symbols. 
The black line is an average at a given $\delta V_g$.
The red line shows the value below which $G_2>e^2/h$ near charge degeneracy ($\delta V_g \approx 0$).
The error bars shown in panels a-d represent the statistical uncertainties (s.e.m.) calculated from 10 successive measurements. 
\normalsize}
\end{figure*}

\begin{figure*}[p]
\renewcommand{\figurename}{\textbf{Extended Data Figure}}
\renewcommand{\thefigure}{\textbf{6}}
\centering\includegraphics [width=1\textwidth]{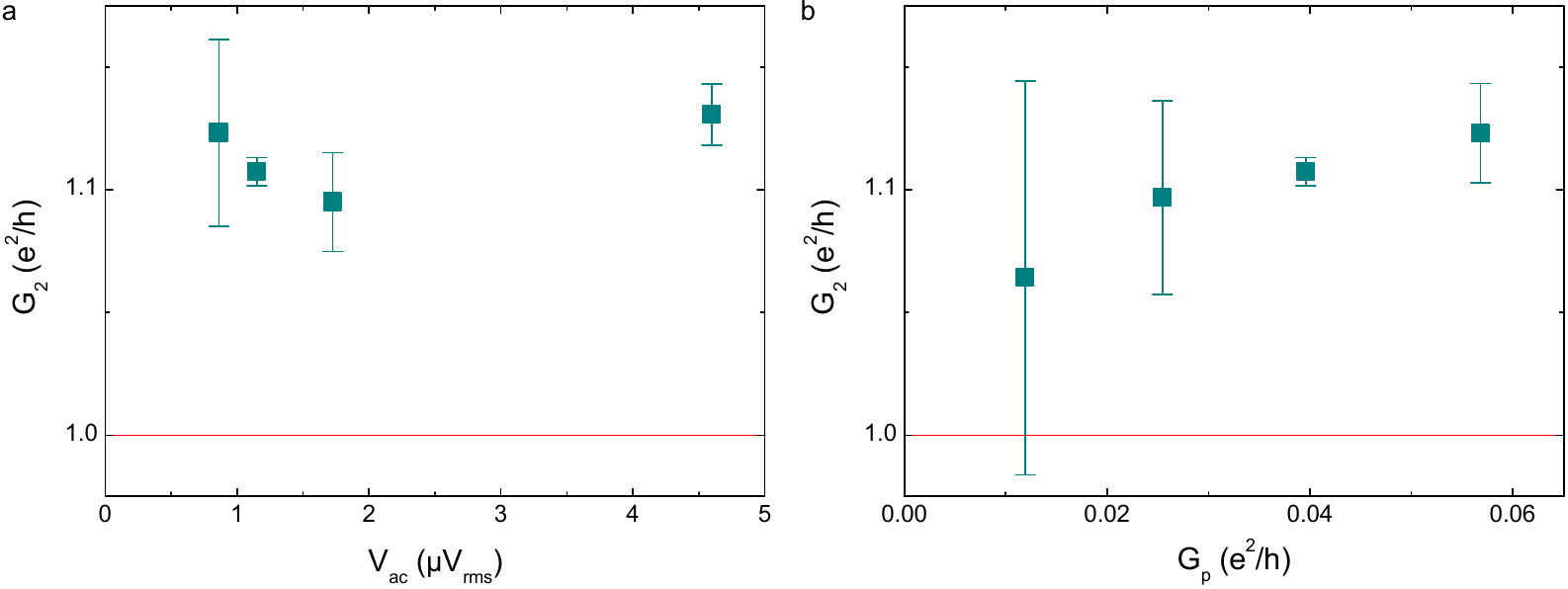}
\caption{\small
\textbf{Robustness to experimental conditions of `in-situ' conductances overstepping $\mathbf e^2/h$.} 
Symbols display the `in-situ' conductance $G_2$ measured at $T\simeq 14$~mK for the QPCs setting $(\tau_1 \simeq 0.77, \tau_2 \simeq 0.93)$, and under different experimental conditions.
We find repeatedly $G_2>e^2/h$.
\textbf{a}, The influence on $G_2$ of the ac injection voltages $V_{1,2,p}$ is explored.
The three lowest $V_\mathrm{ac}$ correspond to $V_1=V_2=V_p=V_\mathrm{ac}$, whereas the fourth data point corresponds to $V_p=V_\mathrm{ac}$ with $V_1=V_2=1.15~\mu\mathrm{V}_\mathrm{rms}$.
\textbf{b},  Exploration of the influence on $G_2$ of the coupling strength of QPC$_p$, characterized by $G_p$.
The error bars represent the statistical uncertainties (s.e.m.) calculated from 20 or more different measurements.
\normalsize}
\end{figure*}
\end{document}